\documentclass[10pt,letterpaper,twocolumn]{article} 

\usepackage{ol}
\usepackage{hyperref}
\usepackage{amsmath}

\begin{document}

\twocolumn[ 

\title{Strong relative intensity squeezing by 4-wave mixing in Rb vapor}

\author{C. F. McCormick, V. Boyer, E. Arimondo$^{*}$, P. D. Lett}

\address{Atomic Physics Division, MS 8424, National Institute of Standards and Technology,
Gaithersburg MD 20899-8424}

\begin{abstract}
We have measured -3.5 dB (-8.1 dB corrected for losses) relative
intensity squeezing between the probe and conjugate beams generated
by stimulated, nondegenerate four-wave mixing in hot rubidium vapor.
Unlike early observations of squeezing in atomic vapors based on
saturation of a two-level system, our scheme uses a resonant
nonlinearity based on ground-state coherences in a three-level
system.  Since this scheme produces narrowband, squeezed light near
an atomic resonance it is of interest for experiments involving cold
atoms or atomic ensembles.
\end{abstract}

\ocis{000.0000, 999.9999.}

] 

Twenty years ago, the pioneering work of Slusher, et al. led to the
first experimental demonstration of a squeezed state of light, using
four-wave mixing (4WM) in sodium vapor\cite{Slusher1985}. Since that
time, many groups have demonstrated squeezing based on 4WM in atomic
vapors under a variety of different conditions \cite{Maeda1987,
Lambrecht1996}

Despite the breadth of these experimental results, 4WM in atomic
vapors has never generated large amounts of squeezing. To our
knowledge, the highest reported level of squeezing from atomic-vapor
4WM is -2.2 dB \cite{Lambrecht1996}, using cold atoms in a cavity.
This is in marked contrast to systems based on 4WM in optical
fibers, for which relative intensity squeezing of up to -4.6 dB
($-10.3$ dB after loss correction) has been observed
\cite{Hirosawa2005}. Systems based on parametric down-conversion
(PDC) have been even more successful, generating better than -9 dB
(measured) of relative intensity squeezing
\cite{Feng2004,Laurat2005}.

It has been suggested that squeezing from 4WM in atomic vapors is
limited by spontaneous emission noise when the nonlinearity is based
on coherences between levels separated by optical transitions
\cite{Slusher1985b}. A large number of recent experiments have
demonstrated the power of atomic-vapor 4WM using nonlinearities
based on ground-state coherences, in which coherent population
trapping and electromagnetically induced transparency (EIT) can
reduce or eliminate spontaneous emission noise. Of particular
relevance to this work, classical noise correlations in such a
$\Lambda$ system were measured in sodium vapor\cite{Grove1997} and
4WM in this system was predicted to generate
squeezing\cite{Shahriar1998,Lukin1998}. Approximately -0.2~dB of
squeezing has been demonstrated in a $\Lambda$ system in rubidium
vapor\cite{vanderWal2003}.

\begin{figure}
\centering
\includegraphics{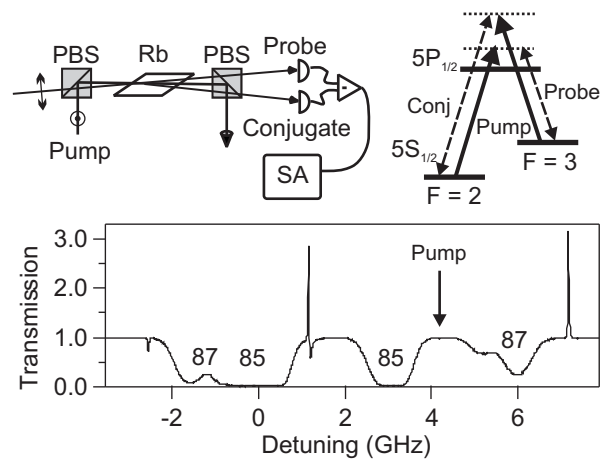}
\caption{TOP LEFT: Experimental schematic;
PBS = polarizing beam splitter, SA = spectrum analyzer.
TOP RIGHT:  Energy-level diagram.
LOWER: Probe transmission versus detuning from
the $^{85}$Rb F = 3 $\rightarrow$ F' transition, for a pump
detuning as indicated by the arrow.} \label{schematic}
\end{figure}

Inspired by these results and motivated by the desire to produce
correlated photons that can interact with laser-cooled atoms to
produce correlated atom-optical beams \cite{Lett2004,Haine2005} we
have revisited the question of squeezing generated by 4WM in atomic
vapors. In this Letter, we report the demonstration of -3.5 dB (-8.1
dB corrected) of relative intensity squeezing between the probe and
conjugate beams produced in a seeded, nearly copropagating,
nondegenerate 4WM scheme in hot rubidium vapor. Our system does not
involve a cavity or cold atoms and has a single-pass net optical
gain of only $\sim 4$. We anticipate that further experimental
improvements to the system could produce significantly greater
squeezing, perhaps making atomic-vapor 4WM competitive with optical
fibers and PDC-based systems as a source of squeezed light.

Our experiment begins with two Ti:Sapphire lasers tuned to the D1
line of rubidium (795 nm). We use 350~mW of light from one as a pump
beam, and up to 0.2~mW of light from the other as a probe. We
combine the pump and probe beams with crossed linear polarizations
and send them into a natural-abundance rubidium vapor cell heated to
125 C (atomic density $\sim3 \times 10^{13}$cm$^{-3}$) with no
magnetic shielding. The beams cross at a small angle ($\sim$0.75
degrees) and the cell is tilted at Brewster's angle ($\sim$8~mm path
length) to minimize reflection losses for the probe polarization.
(See Fig. 1.)  The pump beam waist is 600~$\mu$m while the probe
beam waist is 350~$\mu$m. After the cell the pump beam is filtered
out using a Glan-Taylor polarizer, with a discrimination of
$\sim10^5$:1.

With the pump detuned $\sim$1 GHz to the blue of the $^{85}$Rb F=2
$\rightarrow$ F' transition, we scan the probe by 12 GHz across the
D1 line and observe a number of transmission features (see Fig.
\ref{schematic}). The most prominent are two probe intensity gain
peaks, at the $^{85}$Rb ground-state hyperfine splitting of 3 GHz to
the red and blue of the pump detuning. These are due to forward 4WM
gain.  The probe gain is accompanied by the generation of a
conjugate beam with detuning from the pump opposite to that of the
probe, polarization parallel to the probe, and propagation direction
at the same pump-probe angle but on the opposite side of the pump.
The slight dispersive character of the redder of the two 4WM gain
features is caused by a competition between Raman absorption, EIT,
and the 4WM gain. The absorptive feature at -2.5 GHz is 6.8 GHz to
the red of the pump detuning, and is due to Raman absorption in the
$^{87}$Rb atoms.

\begin{figure}
\centering
\includegraphics{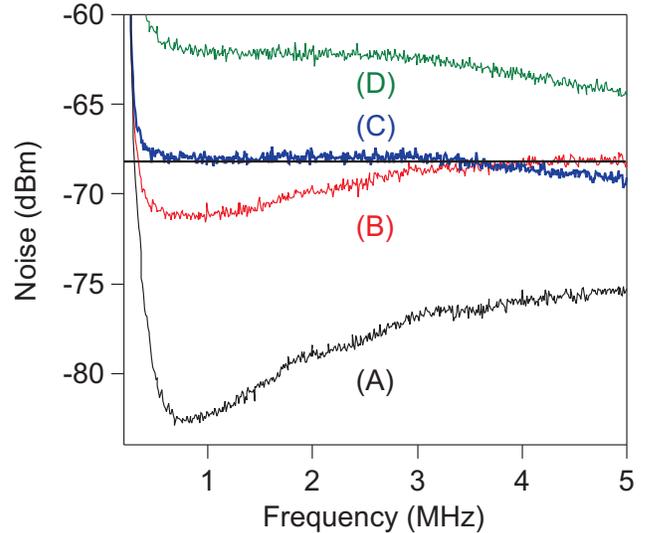}
\caption{(Color online) Relative intensity noise vs. spectrum
analyzer frequency. (A) Electronic noise floor; (B) Observation of
squeezing:  4WM noise for 95 $\mu$W and 85 $\mu$W in probe and
conjugate, respectively; (C) Probe at 180 $\mu$W directed around
cell and through a 50/50 beamsplitter (SQL plus electronic noise);
(D) 4WM, probe beam only with gain, 95 $\mu$W. The solid line at -68
dBm indicates the SQL for 180 $\mu$W in the absence of electronic
noise, inferred from the curve in Fig. \ref{noisevpower}. }
\label{noisevfrequency}
\end{figure}

In order to measure the relative intensity noise between the probe
and conjugate, we switch to a phase-locked probe beam of the same
waist, generated by double-passing a small fraction of the pump
light through a 1.5 GHz acousto-optic modulator. We then calibrate
the standard quantum limit (SQL) ``shot noise" of our system by
picking off the probe before the cell, splitting it with a 50/50
beamsplitter, and directing the resulting beams into a balanced,
amplified photodetector with a transimpedance gain of $10^5$V/A and
82\% quantum efficiency. We measure the spectrum of electrical noise
power of the photodetector output voltage on a spectrum analyzer set
to a 300 kHz resolution bandwidth and a 100 Hz video bandwidth.  The
balanced detection technique subtracts away common-mode noise to
better than 25 dB. The balanced photodetector noise level is a
measure of the SQL for the total amount of optical power arriving at
the photodetector. The shot noise should be independent of
frequency, which is indeed the case within the bandwidth of our
detection electronics, which begins to roll off above 3 MHz. For a
total power of 180 $\mu$W (90 $\mu$W out of each port) the measured
SQL is -68 dBm (see Fig. \ref{noisevfrequency}).

Next, with a pump detuning of 750~MHz and the probe tuned 3.03~GHz
to the red, the light is redirected through the atoms.  Under these
conditions the probe has a gain $g = 4$. As a seeded process the
probe and conjugate should have a power ratio of $g/(g-1)$. In the
absence of the pump, the absorption of the probe is $\sim$ 7\%,
while the absorption at the conjugate detuning is negligible. Due to
this differential absorption the probe and conjugate emerge from the
cell with the same power to within 10\%. We direct these beams onto
the balanced photodetector and measure the noise-power spectrum. For
a total optical power (probe plus conjugate) of 180 $\mu$W, the
noise power is as much as 3.5 dB below the SQL for frequencies from
300~kHz to $\sim$4~MHz (see Fig.~\ref{noisevfrequency}).

\begin{figure}
\centering
\includegraphics{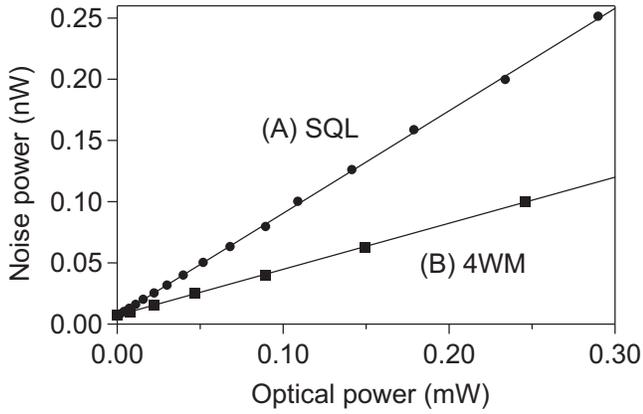}
\caption{Relative intensity noise at 1 MHz for: (A) a 50/50 beamsplit probe
with no atoms, and (B) the 4WM configuration, both vs. total power
falling on the photodetector. Both fits are to straight lines.}
\label{noisevpower}
\end{figure}

To better quantify the degree of squeezing we measure the relative
probe-conjugate noise power at 1~MHz as a function of the total
optical power impinging on the photodetector, and compare it with
the 50/50 beamsplitter SQL measurement as described above, for
various powers (see Fig. \ref{noisevpower}). Fitting both the 4WM
and SQL noise power curves to straight lines, we find that the probe
and conjugate relative intensity noise is 3.5~dB below the SQL.
Using a model for twin-beam losses\cite{Aytur1990,Smithey1992} we
correct for our detector efficiency (82\%) and the transmission of
the optical path from the atoms to the detector (80\%), resulting in
a noise power 8.1$^{+1.4}_{-1.0}$~dB  below the SQL before the exit
window of the cell. This is limited by both the (asymmetric)
absorption in the cell and the imbalance due to the injected probe
light.

We next measure the noise properties of the individual probe and
conjugate beams after the cell, by blocking one and then the other.
The resulting noise spectra are nearly identical, and show increased
noise above the SQL (see Fig. \ref{noisevfrequency}). Since the
probe and conjugate alone have only half the total optical power,
the correct SQL with which to compare them is lower by 3 dB (-71
dBm), since SQL noise power scales linearly with optical power. At 1
MHz the probe and conjugate alone each have 7.6 times more noise
than the SQL. The net probe intensity gain is 4 at the end of the
vapor. If the 4WM system were operating as a perfect
quantum-noise-limited amplifier with this gain (and correcting for
losses) we would expect only 4.9 times the SQL with a
shot-noise-limited input probe.

As a final experimental check, we tune the probe 3.03~GHz to the
blue of the pump, coinciding with the blue 4WM gain peak. For this
configuration the probe and conjugate display relative intensity
noise reduction of only about 2.5~dB. This is consistent with the
fact that the conjugate beam now experiences more absorption than
the probe, leading to less well-balanced optical powers on the
detector.

Our 4WM configuration is related to that discussed by Lukin and
coworkers,\cite{Lukin1998} with probe and conjugate photons playing
complementary roles in the atomic preparation and photon scattering
from the atomic coherence. Previous experiments working with the
$\Lambda$ scheme have tended to operate at low power and near
resonance. The experiments reported here, however, operate in a
different regime and combine a number of features used individually
in previous investigations of $\Lambda$ systems, including a nearly
copropagating geometry, a very large Rabi frequency for the pump
laser ($\Omega_{\mathrm{pump}} \approx 100 \Gamma$), and a large
detuning. Furthermore, the choice of cross-polarization of the pump
and probe reduces the reabsorption, since it leads to optimal EIT on
the D1 line of rubidium by creating a dark state that simultaneously
satisfies multiple $\Lambda$ transitions\cite{Zanon2005}.

The phase stability of the pump and probe beam appears to be an
important but not critical issue; we were able to observe nearly the
same amount of squeezing by using independent pump and probe lasers,
although with a phase-locked pump and probe the spectrum of
squeezing was extended to lower frequencies by about 250~kHz. We
intend to explore further EIT improvements with magnetic shielding
and buffer gases, as well as single-isotope cells of $^{85}$Rb,
which should allow us greater freedom in detuning. In addition, by
pumping on the transitions that we used here as the probe and
conjugate, it should be possible to produce degenerate twin beams
and quadrature squeezing. These photons would be narrowband and at a
frequency required for cold atom experiments.

In conclusion, we have measured -3.5~dB (-8.1~dB corrected for
losses) relative intensity squeezing in forward four-wave-mixing in
hot atomic vapor, in a simple system without a cavity or feedback
loops. This system promises to be an important source of narrowband
squeezed light near atomic transitions.

This work was supported by NASA. CFM was supported by an IC
Postdoctoral Fellowship. We would like to thank Kevin Jones and Luis
Orozco for important discussions.

$^{*}$ Permanent address: Dipartimento di Fisica E. Fermi,
Universit{\'a} di Pisa, Pisa, Italy.

\end{document}